\documentclass[english,apl,manuscript,preprint]{revtex4-1}
\usepackage{ae,aecompl}
\usepackage{beramono}
\usepackage[T1]{fontenc}
\usepackage[latin9]{inputenc}
\usepackage[active]{srcltx}
\usepackage{color}
\usepackage{amsmath}
\usepackage{amssymb}
\usepackage{graphicx}
\usepackage{esint}

\makeatletter

\newcommand{\lyxmathsym}[1]{\ifmmode\begingroup\def\b@ld{bold}
  \text{\ifx\math@version\b@ld\bfseries\fi#1}\endgroup\else#1\fi}

\usepackage[normalem]{ulem}

\usepackage{babel}

\usepackage{xr}
\usepackage{babel}

\makeatother

\usepackage{babel}
\begin{document}

\title{Bulk-impurity induced noise in large-area epitaxial thin films of
topological insulators }

\author{Saurav Islam\textsuperscript{1,{*}}, Semonti Bhattacharyya\textsuperscript{1,{*}},
Abhinav Kandala\textsuperscript{2,3}, \textcolor{black}{Anthony Richardella\textsuperscript{2}},
Nitin Samarth\textsuperscript{2}, Arindam Ghosh\textsuperscript{1,4}}

\address{1. Department of Physics, IISc, Bangalore-560012, India}

\address{2. Department of Physics and Materials Research Institute, The Pennsylvania
State University, University Park, Pennsylvania 16802-6300, USA}

\address{\textcolor{black}{3. IBM T. J. Watson Research Center, Yorktown Heights,
NY 10598, USA}}

\affiliation{4. Centre for Nano Science and Engineering (CeNSE), IISc, Bangalore-560012,
India}

\author{{*}SI and SB contributed equally}
\begin{abstract}
{\normalsize{}We report a detailed study of low-frequency $1/f$-noise
in large-area molecular-beam epitaxy grown thin ($\sim10$\ nm) films
of topological insulators as a function of temperature, gate voltage
and magnetic field. When the fermi energy is within the bulk valence
band, the temperature dependence reveals a clear signature of generation-recombination
noise at the defect states in the bulk band gap. However, when the
fermi energy is tuned to the bulk band gap, the gate voltage dependence
of noise shows that the resistance fluctuations in surface transport
are caused by correlated mobility-number density fluctuations due
to the activated defect states present in the bulk of the topological
insulator crystal with a density $D_{it}=3.2\times10^{17}$\ cm$^{-2}$eV$^{-1}$.
In the presence of magnetic field, noise in these materials follows
a parabolic dependence which is qualitatively similar to mobility
and charge-density fluctuation noise in non-degenerately doped trivial
semiconductors. Our studies reveal that even in thin films of (Bi,Sb)$_{2}$Te$_{3}$
with thickness as low as 10\ nm, the internal bulk defects are the
dominant source of noise. }{\normalsize \par}
\end{abstract}
\maketitle
Topological insulators (TIs) are characterized by gapless linearly
dispersive spin-polarized surface states in the bulk band gap\ \cite{Rev3Qi2011,ZahidHasanRevModPhys.82.3045}.
These materials are promising candidates for various electronic and
spintronic applications\ \cite{Rev4Kong2011} due to the topological
protection of surface states against back scattering from non-magnetic
impurities. However, sensitivity of these surface states to different
types of disorder is still a matter of active debate\ \cite{TiMobility3,TiMobility4,ImpScat1,impurityScatter,Mypaper,Adam2012,Black-Schaffer2012}.
Recently, flicker noise or $1/f$-noise in bulk TI systems has been
established as a powerful tool not only as a performance-marker for
electronic applications but also for defect-spectroscopy and even
band structure determination\ \cite{Mypaper,My_apl,Noise_TI_cascales2015band}.
Noise measurements in exfoliated TI devices from bulk crystals have
revealed that in the thickness range of $50$\ nm to 80\ $\mu$m
charge number fluctuations in the bulk can give rise to the resistance
fluctuations in the surface states\ \cite{Mypaper}. For $10$\ nm
thick exfoliated films however, the underlying disorder dynamics could
not be determined as noise was dominated by universal conductance
fluctuations (UCF), since lateral dimension of this device (800\ nm$\times300$\ nm)
turned out to be of the same order as the phase coherence length ($l_{\varphi}\sim250$\ nm)
itself. However, devices in this thickness range are arguably most
suitable for novel applications\ \cite{Spintronic1,Sacepe2011,SpinPolarizationDetect,NitinTiSpinTorque,Magswitching,proximity_lang2014proximity},
as this thickness is low enough to allow the bottom gate to tune the
chemical potential at both top and bottom surfaces\ \cite{top_bot_PhysRevB.83.241304,SixQPLimit,Kim2012a},
albeit high enough to prevent any overlap between two surface states
which can result in a gap at the Dirac point\ \cite{hybrid_lu2010massive}.
Hence, it is important to evaluate and understand noise in 3D TI devices
with thickness $\sim10$\ nm. In this work, we have studied large
area (1\ mm$\times$0.5\ mm) TI films with $10$\ nm thickness,
where mesoscopic conductance fluctuations are averaged out and underlying
defect fluctuations can be detected.

\begin{figure}
\includegraphics[bb=0bp 10bp 226bp 96bp]{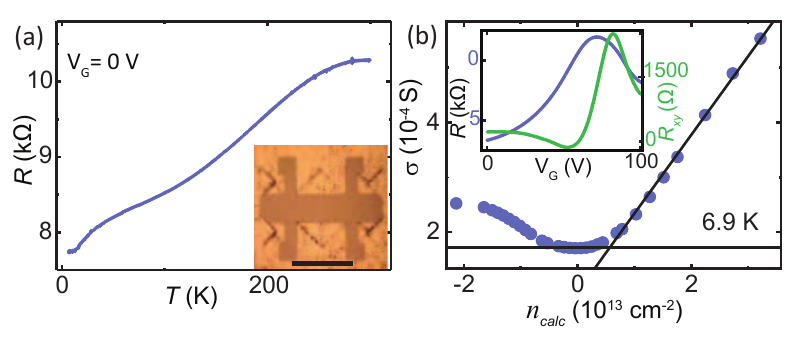}

\protect\caption{(a) Temperature ($T$) dependence of resistance ($R$) at gate voltage
$V_{G}=0$\ V in device BST1. The inset shows optical image of the
device BST1. The black-line shows a scale bar with length 1\ mm.
(b) Conductance ($\sigma=\frac{L}{RW}$) vs. calculated number density
($n_{calc}=\frac{C_{STO}\left(V_{G}-V_{D}\right)}{e}$) at 6.9\ K.
The black lines are fit of this data according to the equations \ref{eq:coulomb imp 1}
and \ref{eq:coulomb imp 2}. Inset shows $V_{G}$ dependence of longitudinal
($R$) and transverse ($R_{xy}$) resistance in device BST1 at 6.9\ K.
$R_{xy}$ was obtained at magnetic field $B=-0.5$\ T.\label{fig:(a)-basic characteristics}}
\end{figure}

The devices studied in this paper were fabricated from thin (thickness,
$d=10$\ nm) films of (Bi,Sb)$_{2}$Te$_{3}$ (BST) grown by molecular
beam epitaxy on $\left\langle 111\right\rangle $ SrTiO$_{3}$ (STO)
substrates with a metallic back-coating of Indium that is used as
back gate electrode\ \cite{eeFMTI1_kandala2013growth}. All the measurements,
except temperature ($T$) dependence of $1/f$-noise, were performed
in device BST1 (gate voltage at charge neutrality point $V_{D}=66$\ V
and 71\ V at two consecutive thermal cycles).
The $T$-dependence of noise was performed in BST2 ($V_{D}>150$\ V).
Both devices were intrinsically hole-doped. Hall bar geometry with
a channel length of 1\ mm and width 0.5\ mm was defined by mechanical
patterning (inset, Fig.\ \ref{fig:(a)-basic characteristics}a).
While, non-invasive hall bar geometry minimizes contribution of contacts
in electrical transport and noise measurements\ \cite{karnatak2016current,Ghatak2014},
the mechanical etching process prevents the film from being exposed
to chemicals (e-beam resist and solvents), resulting in a superior
surface quality. The large dielectric constant of the STO substrate
at cryogenic temperatures allows effective electrical back gating
for tuning the chemical potential in the sample\ \cite{STO_TI4tian2014quantum,STO_TI3:/content/aip/journal/apl/101/12/10.1063/1.4754108,Kim2012a}.
Electrical transport and noise measurements were conducted down to
$T=5$\ K and up to magnetic field $(B)$ of 13\ T.

The decrease in resistance ($R$) with decreasing $T$ (Fig.\ \ref{fig:(a)-basic characteristics}a)
indicates metallic behavior, which is similar to the $R$-$T$ behavior
reported previously in similar thin exfoliated TI devices with negligible
bulk conduction\ \cite{Mypaper}. However, in this case, the measured
intrinsic Hall number density is $n_{Hall}=0.9\times10^{14}$\ cm$^{-2}$
(at $T=6.9$\ K and gate voltage $V_{G}=0$\ V), which is high compared
to typical number density of surface charge carriers in TI\ \cite{STO_TI4tian2014quantum},
indicating that the Fermi energy is located within the valence band,
and holes in the bulk of the material are mainly responsible for such
metallic behavior. This intrinsic high doping leads to a high value
of the gate voltage ($V_{G}=66$\ V), at charge neutrality point,
which is identified by a maximum in the $R$ and a change of sign
in transverse resistance $R_{xy}$ (inset, Fig.\ \ref{fig:(a)-basic characteristics}b).
In order to evaluate the extent of disorder, we have fitted $\sigma$-$n_{calc}$
data (Fig\ \ref{fig:(a)-basic characteristics}b) (where $\sigma=\frac{L}{RW}$
and $n_{calc}=\frac{C_{STO}\left(V_{G}-V_{D}\right)}{e}$ within the framework
of charge-impurity limited scattering of Dirac fermions\ \cite{sigma_nculcer2010two,Kim2012a},
so that

\begin{eqnarray}
\sigma\sim E\left|\frac{n}{n_{i}}\right|\left[e^{2}/h\right]\ \  & \mathrm{for}\ \  & n>n^{*}\label{eq:coulomb imp 1}
\end{eqnarray}

and

\begin{eqnarray}
\sigma\sim E\left|\frac{n^{*}}{n_{i}}\right|\left[e^{2}/h\right]\ \  & \mathrm{for}\ \  & n<n^{*}\label{eq:coulomb imp 2}
\end{eqnarray}
where $n^{*}$ is the residual carrier density in electron and hole
puddles, and $E$ is a constant that depends on the Wigner\textendash Seitz
radius $r_{s}$. The extracted value of number density of Coulomb
trap is $n_{i}=1\times10^{14}$ cm$^{-2}$, which is quite high compared
to typical density of Coulomb traps at oxide-channel interface in
graphene-FETs\ \cite{chen2008charged}, but matches well with the
number density of charged disorders in TI-devices\ \cite{Kim2012a},
and can be attributed to the bulk charged-defects in TI.

\begin{figure*}
\includegraphics{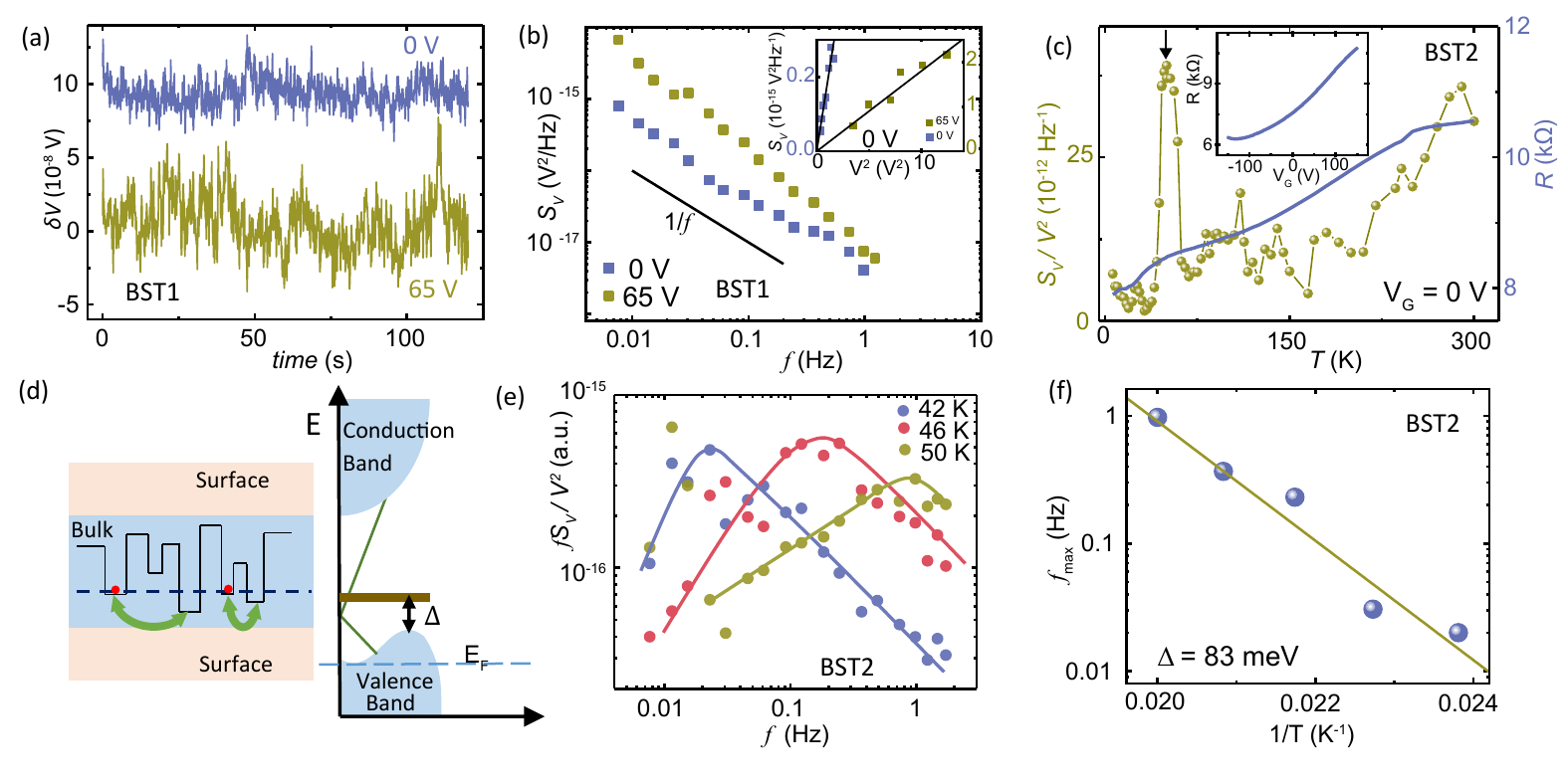}

\protect\caption{(a) Voltage fluctuations ($\delta V$) in time domain in BST1 at 6.9\ K
and $V_{G}=0$\ V and 65\ V obtained with fixed biasing current.
(b) Typical voltage power spectral density ($S_{V}$) obtained from
time dependent fluctuations shown in (a), indicating 1/$f$ type characteristics.
Inset shows $S_{V}$ at 1\ Hz for as a function of $V^{2}$ showing
linear characteristics at the same values of $V_{G}$. (c) Normalized
voltage power spectral density ($S_{V}/V^{2}$) and resistance ($R$)
as a function of $T$ at $V_{G}=0$\ V in device BST2. The arrow
shows the peak at 50\ K. The inset shows $V_{G}$ dependence of $R$
in this device at 6.9\ K. (d) Schematic showing two different noise
mechanisms. Left panel: trapping-detrapping of charge carriers in
the bulk of the material. Right panel: generation-recombination of
charge-carriers between valence band and defect state. (e) Frequency
($f$) and voltage ($V$) normalized power spectral density ($fS_{V}/V^{2}$)
as a function of frequency ($f$) in device BST2. Traces at three
temperatures are shown here for clarity. The solid lines are guide
to the eye. (f) Corner frequencies of power spectral density ($f_{max}$)
as a function of $1/T$ extracted from (e). The solid line shows fit
to this data using the eq. $f_{max}=f_{0}exp(-\Delta E/k_{B}T)$\ \cite{JamalDeen}(here
$\Delta E,$ $k_{B}$ and $f_{0}$ are the interband transition energy,
Boltzmann constant and phonon frequency scales, respectively).\label{fig:noise basic}}
\end{figure*}

An AC four-probe Wheatstone bridge technique was used for noise measurements\ \cite{scofield1987ac,SaquibPRB}.
The voltage fluctuations as shown in Fig.\ \ref{fig:noise basic}a
were recorded as a function of time using a 16-bit digitizer. The
time-series data were then digitally processed to obtain the power-spectral-density
(PSD, $S_{V}$) as a function of frequency ($f$) (Fig\ \ref{fig:noise basic}b).
In both devices, we found $S_{V}\propto1/f^{\alpha}$, where the frequency
exponent $\alpha\thickapprox1\lyxmathsym{\textendash}1.2$. The $S_{V}$
shows a quadratic dependence with bias ($V$) (inset, Fig.\ \ref{fig:noise basic}b),
which ensures that we are in the Ohmic regime.

In device BST2, the chemical potential is intrinsically located within
the valence band and only the bulk electronic states contribute to
transport ($R$ changes by less than $50\%$ over the entire gate
voltage range of $\pm100\ V$ (inset, Fig.\ \ref{fig:noise basic}c)).
The magnitude of noise ($\frac{S_{V}}{V^{2}}$ at $f=1$\ Hz) (Fig.\ \ref{fig:noise basic}c)
shows a gradual increase with increasing $T$ along with a strong
peak at $T=50$\ K. The gradual increase in noise can be associated
with thermally activated defect dynamics in metallic diffusive systems\ \cite{activated_raychaudhuri2002flicker}.
The strong peak in noise magnitude at $T=50$\ K bears close resemblance
to the noise maximum associated with bulk generation-recombination
(G-R) process, which was reported recently for mechanically exfoliated
heavily-doped TI devices \cite{Mypaper}. The process is schematically
explained in Fig.\ \ref{fig:noise basic}d, which is a characteristic
feature of Bi-chalcogenide-based TI devices that depends on the energetics
of the impurity bands. In order to confirm the G-R mechanism for the
noise peak at $T=50$\ K, we note the timescale associated with this
process ($\tau$) gives rise to a frequency maximum ($f_{max}=\frac{1}{2\pi\tau}$)
in the power spectral density (PSD) at temperatures close to the peak
(Fig.\ \ref{fig:noise basic}e). The value of the energy gap associated,
as obtained from the fitting of activated behavior of $f_{max}$ (Fig.\ \ref{fig:noise basic}f)
as a function of $T$ is $\Delta E\sim83$\ meV, which indicates
that the impurity band responsible for this noise is located at $\sim83$\ meV
above the bulk valence band 

\begin{figure}
\includegraphics[bb=0bp 10bp 227bp 416bp]{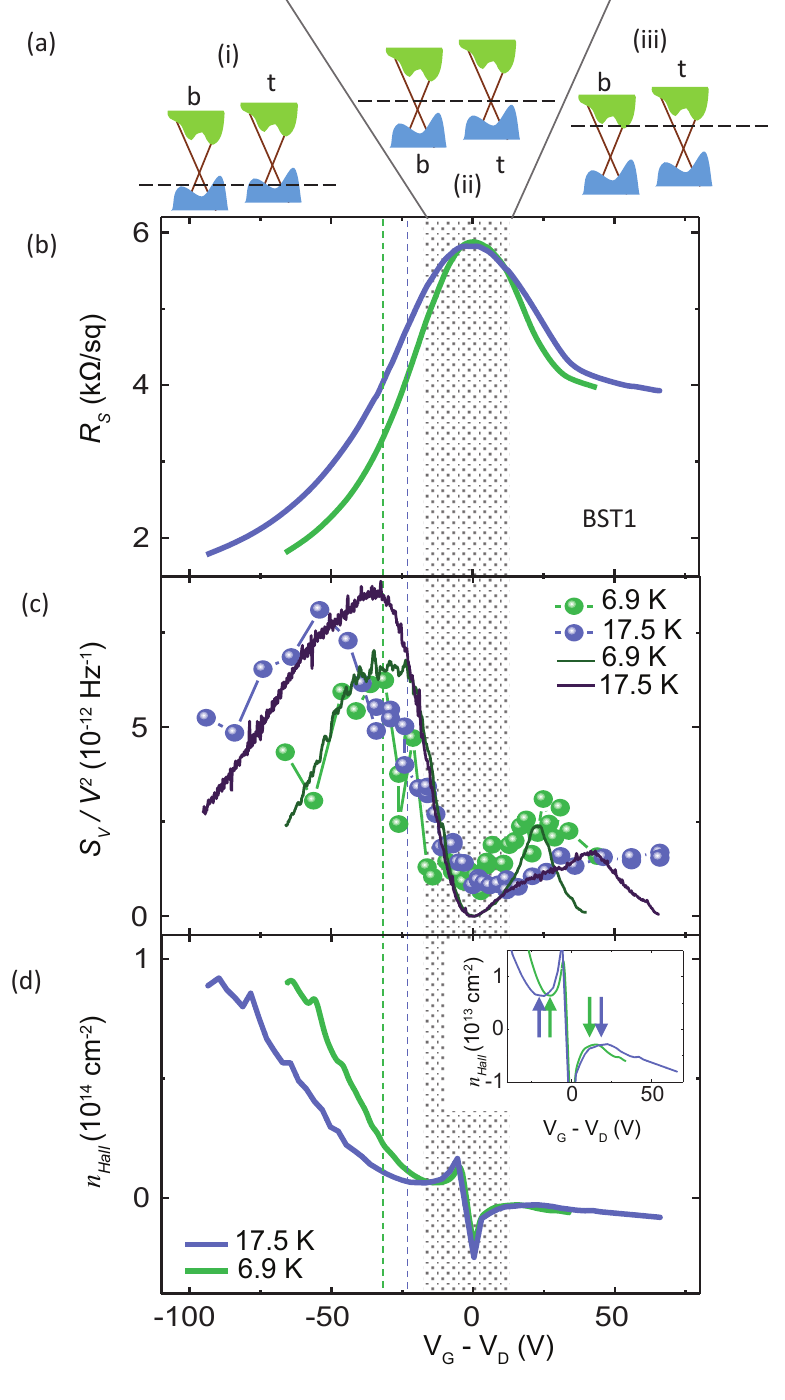}

\protect\caption{(a) Schematic showing Fermi energy at top (t) and bottom (b) surfaces
of TI film at different ranges of $V_{G}$. Region (i) indicates dominant
hole contribution from surface and bulk states. Region (ii) indicates
the charge puddle dominated regime close to the Dirac points for both
the surfaces. Region (iii) indicates the surface-electron dominated
regime for both the surfaces. (b) Sheet-resistance ($R_{S}=\frac{WR}{L}$,
where $W$, $L$ and $R$ are width, length and resistance of the
channel respectively), (c) Normalized voltage power spectral density
$\left(\frac{S_{V}}{V^{2}}\right)$ and (d) Hall number density of
charge carriers ($n_{Hall}$) as a function of effective gate voltage
($V_{G}-V_{D}$) at $6.9$\ K and $17.5$\ K. Inset of (d) highlights
into the low number-density area, the arrows highlight the extrema.
$n_{Hall}$ at 17.5\ K was extracted from extrapolating the data
at 6.9\ K, which was experimentally measured (see section\ 8 in
Supplementary information). The dashed lines indicate the onset of
bulk conduction where $n_{Hall}=10^{13}$\ cm$^{-2}$\label{fig:Normalized-voltage-power}.}
\end{figure}

The $V_{G}$-dependence of sheet-resistance ($R_{S}$) shows graphene-like\ \cite{atinapl,AtinACSNano,atin_apl_cvd}
ambipolar transport at both 6.9\ K and 17.5\ K (Fig.\ \ref{fig:Normalized-voltage-power}b).
The increase of $V_{D}$ at the higher $T$ is caused by the reduction
of dielectric constant of STO\ \cite{hong2011ultra}. Voltage normalized power spectral density
$\left(\frac{S_{V}}{V^{2}}\right)$, \textit{i.e.}, the magnitude
of noise is shown as a function of effective gate voltage $\left(V_{G}-V_{D}\right)$
at 6.9\ K and 17.5\ K (Fig.\ \ref{fig:Normalized-voltage-power}c).
It is minimum close to the Dirac point $\left(V_{G}-V_{D}=0\right)$,
but increases rapidly ($\sim10$ times) with increasing $\left|V_{G}-V_{D}\right|$,
and reaches a maximum (at $\left(V_{G}-V_{D}\right)=-33$\ V, at
6.9\ K). With further increase in $\left|V_{G}-V_{D}\right|$, noise
magnitude reduces.

To correlate this behavior with the nature and sign of the charge
carrying species, we have measured the Hall number density ($n_{Hall}$)
as a function of $\left(V_{G}-V_{D}\right)$ at 6.9\ K. Change in
the back-gate capacitance was estimated from the $R$-$V_{G}$ data,
and used to extract the $V_{G}$ dependence of $n_{Hall}$ at 17.5\ K.
Fig.\ \ref{fig:Normalized-voltage-power}d shows that the back-gate
can tune the state of doping from p-type to n-type over the entire
device. Our data clearly illustrates that the noise magnitude is enhanced
for $\left|n_{Hall}\right|\gtrsim10^{13}$\ cm$^{-2}$ on the hole
side, which is the number density associated with the edge of the
bulk valence band\ \cite{STO_TI4tian2014quantum,Kim2012a}, demonstrating
increased impact of disorder on the bulk charge transport as expected.
Due to non-linear dependence of capacitance on $V_{G}$\ \cite{STO_TI4tian2014quantum},
the number density in the accessible range of $V_{G}$ did not increase
beyond $n_{Hall}=-0.8\times10^{12}$\ cm$^{-2}$ in the electron-side,
and hence the noise magnitude saturated to a lower value.

In Fig.\ \ref{fig:Normalized-voltage-power}d the two extrema at
the number density (\textit{e.g.} $-13$\ V and 15.7\ V, at 6.9\ K)
indicate the transition from pseudo-diffusive (shaded area) to diffusive
electrical transport\ \cite{STO_TI3:/content/aip/journal/apl/101/12/10.1063/1.4754108,STO_TI4tian2014quantum}.
However, unlike mesoscopic devices of graphene\ \cite{AtinACSNano,Rossipaper1}
or TIs\ \cite{Mypaper}, the noise does not show any specific signature
at these transition points in our device as mesoscopic effects average
out in the large area thin films where $L,W\gg l_{\varphi}$ ($L$
and $W$ are length and width of the channel).

For a more quantitative understanding, we have fitted the $V_{G}$-dependence
of $\frac{S_{V}}{V^{2}}$ data (Fig\ \ref{fig:Normalized-voltage-power}c)
using the framework of correlated mobility-number density fluctuations
model\ \cite{Jayaraman1989}. According to this model, the noise
in the channel is affected by the trapping-detrapping of charges in
gate dielectric-channel interface (Fig.\ \ref{fig:noise basic}d).
The main components of the noise are,

\noindent 1. The fluctuations of the number of charge carriers in
the channel due to tunneling to and from the oxide traps at the channel-gate
dielectric interface.

\noindent 2. The mobility fluctuations caused by the Coulomb scattering
induced by fluctuating trapped charges.

\noindent The total noise can be expressed as, 
\begin{equation}
\frac{S_{V}}{V^{2}}=\frac{D_{it}k_{B}T}{dWL}\left(\frac{d\sigma}{dn}\right)^{2}\left(\frac{J_{1}}{\sigma^{2}}+\frac{J_{2}}{\sigma}+J_{3}\right)\label{eq:jayaraman}
\end{equation}

where $J_{1}=\frac{1}{8\alpha}$ represents a pure number fluctuation,
$J_{3}=\int A^{2}(x)\frac{\tau_{_{T}}}{1+\left(2\pi f\tau_{_{T}}\right)^{2}}dx$
represents pure mobility fluctuations and $J_{2}=\int2A(x)\frac{\tau_{_{T}}}{1+\left(2\pi f\tau_{_{T}}\right)^{2}}dx$
represents combined number and mobility fluctuations ($\alpha$ is
the decay constant for the spatially decaying time constant $\tau_{_{T}}$
of a typical trapping event and $A(x)$ is the scattering constant)
and can be evaluated using phenomenological values\ \cite{Jayaraman1989}.
$D_{it}$, $k_{B}$, $W$, $L$, $\sigma$, $n$, $x$ are the areal
trapped charge density per unit energy, Boltzmann constant, width
of the channel, length of the channel, conductance and number density
of charge carriers, axis in the direction perpendicular to the channel
respectively, $f=1$\ Hz frequency and $d=1$\ nm is the distance
over which the tunneling is effective. The data at 6.9\ K and 17.5\ K
were fitted independently to extract the values of $D_{it}$. We found,
$D_{it}=3.2\times10^{17}$\ cm$^{-2}$eV$^{-1}$ at 6.9\ K, which
is several orders of magnitude higher compared to typical trap density
at oxide surface\ \cite{Jayaraman1989,paul2016percolative}. However,
considering an energy window of $k_{B}T$, the number of activated
trapped states, $n_{I}=1.3\times10^{14}$ cm$^{-2}$, which matches
with the number density of Coulomb traps $n_{i}=1\times10^{14}$ cm$^{-2}$,
extracted from the fitting of the $\sigma$-$n$ (Fig.\ \ref{fig:(a)-basic characteristics}b)
showing that the extracted value of $D_{it}$ is not an artefact and
hence this high density of charge impurities can be attributed to
intrinsic bulk defect states in (Bi,Sb)$_{2}$Te$_{3}$\ \cite{Kim2012a}.
Similar $D_{it}$ ($8\times10^{17}$\ cm$^{-2}$eV$^{-1}$) was found
for 17.5\ K. The origin of the slight elevation of noise magnitude
at 17.5\ K can be attributed to the thermal activation of the charge-defects.

\begin{figure}[b]
\includegraphics[bb=0bp 10bp 227bp 274bp]{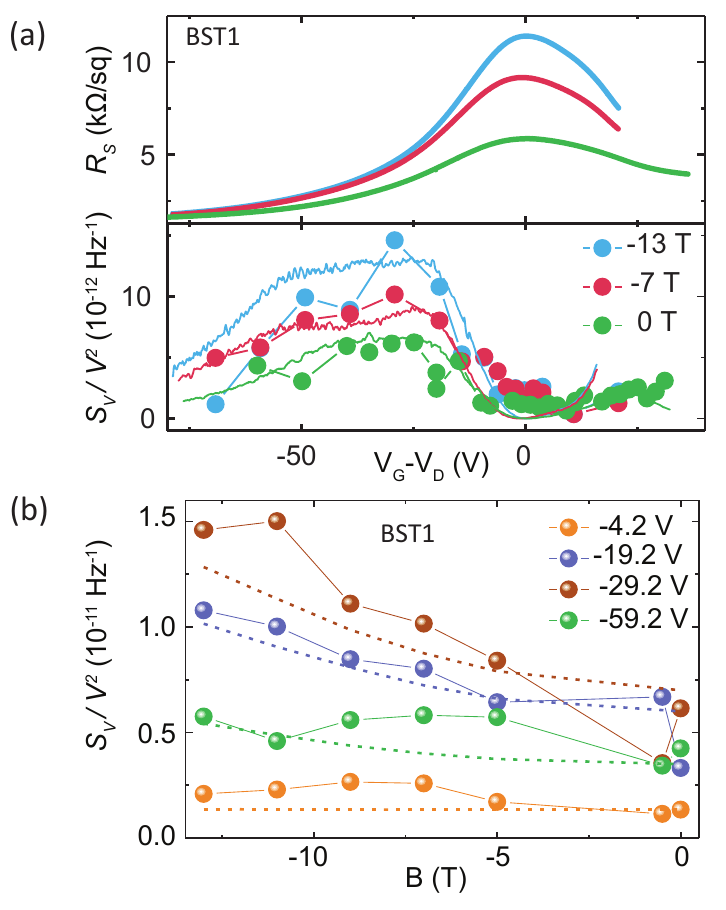}

\protect\caption{(a) $R_{S}$ and $S_{V}/V^{2}$ as a function of $(V_{G}-V_{D})$
at magnetic field values from $0$\ T to $-13$\ T at 6.9\ K. The
continuous lines in the lower panel show the fit of this data according
to the eq.\ \ref{eq:B dependence}. (b) Magnetic field ($B$) dependence
of $S_{V}/V^{2}$ at different values of $(V_{G}-V_{D})$. The dashed
lines show fitting according to the equation\ \ref{fig: magnetic field dependence}.\label{fig: magnetic field dependence}}
\end{figure}

We have performed magnetic field ($B$) dependent measurement of noise
in our devices to verify the correlated charge and mobility fluctuation-related
origin of noise. In the entire range of $V_{G}$ ($\left(V_{G}-V_{D}\right)=-69$\ V
to 21\ V) the noise follows a parabolic dependence with magnetic
field (Fig.\ \ref{fig: magnetic field dependence}b) which also reflects
in the $V_{G}$-dependence of noise at different values of magnetic
field (Fig.\ \ref{fig: magnetic field dependence}a). Drude theory
of metals predicts this type of behavior for a charge-fluctuation
or mobility fluctuation type noise in a non-degenerately doped semiconductor
with $\mu_{H}B\ll1$ ( where $\mu_{H}\sim10$\ cm$^{2}$/V.s is the
Hall mobility)\ \cite{Noise_B1_kleinpenning19801,Noise_B2_voorde1981magnetic}.
Fig.\ \ref{fig: magnetic field dependence}a and b shows fit of this
data according to the eq.

\begin{equation}
\frac{S_{V}}{V^{2}}(B)=\frac{S_{V}}{V^{2}}(0)\left(1+\left(\beta\mu_{H}B\right)^{2}\right)\label{eq:B dependence}
\end{equation}
Here, $\beta$ is a fitting parameter. Although $\mu_{H}B\ll1$ in
our devices, the extracted value of $\beta=7.5$ is quite high compared
to the theoretical value expected for charge or mobility fluctuation
noise\ \cite{Noise_B2_voorde1981magnetic}. In the absence of a theory
which accounts for the magnetic field dependence of correlated mobility-number
density fluctuation noise for a doped semiconductor, we cannot comment
whether this strong dependence on magnetic field results from simple
geometric effects or is a consequence of the topological nature of
the charge-carriers. However, we note that the behavior does not change
as the chemical potential is tuned through surface and bulk bands
of topological insulators.

The results of our experiments clearly indicate that even in thin
films of topological insulator (BiSb)$_{2}$Te$_{3}$ with thickness
as low as $\sim10$\ nm, the bulk chalcogenide defects are mainly
responsible for resistance fluctuations for both surface and bulk
electronic states. Whereas for the bulk valence band, generation-recombination
processes in between the valence band and the impurity states are
the main reason of resistance fluctuations, the noise at the surface
states are caused by the correlated mobility-charge density fluctuation
caused by trapping-detrapping of surface charge carriers by the charged-defects
in the bulk of the crystal. This scenario is different from our earlier
study of noise in thicker exfoliated topological insulator devices\ \cite{Mypaper}
where generation recombination of charge carriers at the bulk gives
rise to a fluctuation of potential landscape at the surface which
results in a mobility fluctuation noise. Further theoretical understanding
is needed for the strong magnetic field dependence of noise in topological
insulators.

AG, SB and SI acknowledge support from DST, India. AR and NS acknowledge
support from The Pennsylvania State University Two-Dimensional Crystal
Consortium \textendash{} Materials Innovation Platform (2DCC-MIP)
which is supported by NSF cooperative agreement DMR-1539916.

\bibliographystyle{ieeetr}

\newpage.

\end{document}